\documentclass[%
 reprint,
superscriptaddress,
 amsmath,amssymb,
 aps,
prb,
]{revtex4-1}

\usepackage{graphicx}
\usepackage{theorem}
\usepackage{dcolumn}
\usepackage{longtable}
\usepackage{bm}
\usepackage{color} 

\newcommand{\SSS}{\scriptscriptstyle}

\newcommand{\ie}{{\textit{i.e.}}}
\newcommand{\eg}{{\textit{e.g.}}}

\newcommand{\etc}{{\textit{etc.}}}

\newcommand{\muB}{{\mu_{\SSS\text{B}}}}

\makeatletter
\renewcommand{\figurename}{Fig.}
\renewcommand{\tablename}{Table}
\renewcommand{\fnum@figure}[1]{\textbf{\figurename~\thefigure} }
\renewcommand{\fnum@table}[1]{\textbf{\tablename~\thetable} }
\makeatother
\renewcommand{~}{\,}

\begin{document}

\preprint{APS/123-QED}

\title{Electron spin coherence on a solid neon surface}

\author{Qianfan Chen}

\email{qianfan.chen@anl.gov}
\affiliation{Center for Nanoscale Materials, Argonne National Laboratory, Argonne, Illinois 60439, USA }

\author{Ivar Martin}
\affiliation{ Materials Science Division, Argonne National Laboratory, Argonne, Illinois 08540, USA}

\author{Liang Jiang}
\affiliation{Pritzker School of Molecular Engineering, University of Chicago, Chicago, Illinois 60637, USA }

\author{Dafei Jin}

\email{djin@anl.gov}
\affiliation{Center for Nanoscale Materials, Argonne National Laboratory, Argonne, Illinois 60439, USA }

\date{\today}

\begin{abstract}
A single electron floating on the surface of a condensed noble-gas liquid or solid can act as a spin qubit with ultralong coherence time, thanks to the extraordinary purity of such systems. Previous studies suggest that the electron spin coherence time on a superfluid helium (He) surface can exceed 100~s. In this paper, we present theoretical studies of the electron spin coherence on a solid neon (Ne) surface, motivated by our recent experimental realization of single-electron charge qubit on solid Ne. The major spin decoherence mechanisms investigated include the fluctuating Ne diamagnetic susceptibility due to thermal phonons, the fluctuating thermal current in normal metal electrodes, and the quasi-statically fluctuating nuclear spins of the $^{21}$Ne ensemble. We find that at a typical experimental temperature about 10~mK in a fully superconducting device, the electron spin decoherence is dominated by the third mechanism via electron-nuclear spin-spin interaction. For natural Ne with 2700~ppm abundance of $^{21}$Ne, the estimated inhomogeneous dephasing time $T_{2}^{*}$ is around 0.16~ms, already better than most semiconductor quantum-dot spin qubits. For commercially available, isotopically purified Ne with 1~ppm of $^{21}$Ne, $T_{2}^{*}$ can be $0.43$~s. Under the application of Hahn echoes, the coherence time $T_{2}$ can be improved to $30$~ms for natural Ne and $81$~s for purified Ne. Therefore, the single-electron spin qubits on solid Ne can serve as promising new spin qubits.
\end{abstract}

\maketitle
\pretolerance=8000 


\section{\label{sec:level1}Introduction}

Qubits are the simplest building blocks of quantum information systems. Among the various forms of qubits today, electron spin qubits are particularly promising owing to their comparatively long coherence time and fast operation speed~\cite{herbschleb2019,ye2019,burkard2021}. Coherence time, which describes how long a quantum superposition persists between the $|0\rangle$ and $|1\rangle$ states, is arguably the most important measure of a qubit's performance.

By convention, an experimentally measured coherence time is denoted by $T_2^*$, which consists of a homogeneous contribution labeled by $T_2$ and an inhomogeneous contribution labeled by  $T_2'$. They are related via~\cite{chavhan2009}
\begin{equation}
\frac{1}{T_2^*} = \frac{1}{T_2} + \frac{1}{T_2'}.
\end{equation}
The inhomogeneous contribution reflects the dephasing and spectral broadening from extrinsic mechanisms, such as the spatiotemporal variation of local environment or measurement apparatus. These are practically unavoidable but can be mitigated by various dynamical decoupling techniques, for instance, the Carr-Purcell-Meiboom-Gill (CPMG) pulse sequences~\cite{yang2017}. The homogeneous contribution is more intrinsic and can be further separated as~\cite{abragam1983,slichter1996,yang2017}
\begin{equation}
\frac{1}{T_2} = \frac{1}{2T_1} + \frac{1}{T_\varphi},
\end{equation}
with $T_1$ the relaxation time associated with irreversible processes and $T_\varphi$ the pure dephasing time from elastic time-varying processes~\cite{yang2017,burkard2021}.

Today, the most common electron spin qubits are nano-fabricated, gate-defined quantum dots (QDs) in semiconductor heterojunctions or semiconductor-oxide interfaces~\cite {petta2005,koppens2006,hanson2007,zwanenburg2013,vandersypen2017,leon2021}. The measured $T_2^*$ in these systems ranges from several nanoseconds to hundreds of microseconds~\cite{stano2021}. With the CPMG pulse sequences that can partially cancel the inhomogeneous dephasing, the extended coherence time $T_2^{\text{CPMG}}$ can be much longer than the original  $T_2^*$~\cite{stano2021}. For example, the state-of-the-art electron spin qubits have $T^{*}_{2}\simeq 20~\mu$s and $T_2^{\text{CPMG}}\simeq 100~\mu$s at a Si/SiGe interface~\cite{yoneda2018}, and $T^{*}_{2}\simeq 120~\mu$s and $T_2^{\text{CPMG}}\simeq 28$~ms at a Si/SiO$_{2}$ interface~\cite{veldhorst2014}.

In addition to the gate-defined semiconductor QD qubits, there are other types of electron spin qubits that are bound to ions or molecules and have even longer coherence time. For example, the homogeneous coherence time $T_{2}$ of an electron in a rare-earth ion (REI) can be between $50~\mu$s and $23$~ms in an erbium doped calcium tungsten oxide crystal Er$^{3+}$:CaWO$_{4}$~\cite{bertaina2007,dantec2021}, and on the order of  several milliseconds in a magnetic molecule~\cite{zadrozny2015}. Furthermore, the CPMG extended coherence time $T_{2}^{\text{CPMG}} $ (close to $T_2$) of an electron on a phosphorus donor in an isotopically purified silicon (Si) layer can reach about 0.5~s~\cite{muhonen2014}. Nonetheless, these electron spin qubits, due to the yet indeterministic site occupation, confront challenges to scale up at this time. In this circumstance, novel electron spin qubits bearing both ultralong coherence time approaching or exceeding 1~s, and deterministic positions like the gate-defined QD qubits have, will provide compelling advantages in building scalable quantum information systems.

Noble-gas elements helium (He) and neon (Ne) are the two most stable elements in nature. At low temperatures, their condensed, liquid and solid phases with vanishing impurities, small dielectric permittivity, and tiny concentration of spinful isotopes ($^3$He and $^{21}$Ne) can serve as ultraclean low-noise environments for any particulate qubits (electrons, ions, atoms, \etc)~\cite{platzman1999,dahm2002,dykman2003,Smolyaninov2000,Smolyaninov2001,jin2020} When an excess electron approaches a condensed liquid He or solid Ne from vacuum, it can form surface states under two effects: (a)  a repulsive barrier on the order of 1~eV due to the Pauli exclusion between the excess electron and atomic shell electrons; (b) an attractive potential from the image charge inside the liquid or solid due to the polarization by the electron~\cite{cole1969,cole1971,leiderer1992}. In the past two decades, there have been considerable efforts in utilizing the motional (charge) or spin states of a single electron on a liquid (superfluid) He surface to build a qubit~\cite{platzman1999,dahm2002,dykman2003,papageorgiou2005,lyon2006,sabouret2008,schuster2010,yang2016,koolstra2019}. The spin coherence time is predicted to be in excess of 100~s~\cite{lyon2006}. However, due to the practical challenges in suppressing liquid-surface vibration and performing single-spin readout, electron qubits on liquid He, whether as charge or spin qubits, have not been achieved so far.

In our recent experiment, we realized an electron motional (charge) qubit on a solid Ne surface~\cite{zhou2022}, as the first demonstration of electron qubits in condensed noble-gas systems. At our experimental temperature around 10~mK, the electron's out-of-plane motion in $z$ is frozen on the ground state. Its in-plane motion in $xy$ is confined by the fabricated on-chip electrodes and shows a trapping lifetime exceeding two months. The qubit states are defined as the two lowest motional states in $y$ with the same ground state in $x$~\cite{zhou2022}. We achieved strong coupling between a single electron and a single microwave photon in an on-chip superconducting resonator and measured the energy relaxation time $T_{1}$ of 15~$\mu$s, and phase coherence time $T_{2}$ over 200~ns~\cite{zhou2022}. Although such performance is already state-of-the-art compared with the existing semiconductor charge qubits, the motional coherence time is not yet long. This may be induced by the background charge noise due to the motion of remnant electrons inside the resonator or a time-varying trapping potential due to the motion of Ne atoms on an imperfect (presumably rough and porous) surface~\cite{zhou2022}. To overcome this deficiency, we have started to develop electron spin qubits in this system, which can potentially yield spin coherence time over 1~s.

The purpose of this paper is to theoretically investigate the major decoherence mechanisms of an electron spin on a solid Ne surface and calculate the coherence time. This study has not been systemically carried out before.

\section{Decoherence mechanisms}

\begin{figure}[htb]
	\begin{center}
		\includegraphics[scale=0.5]{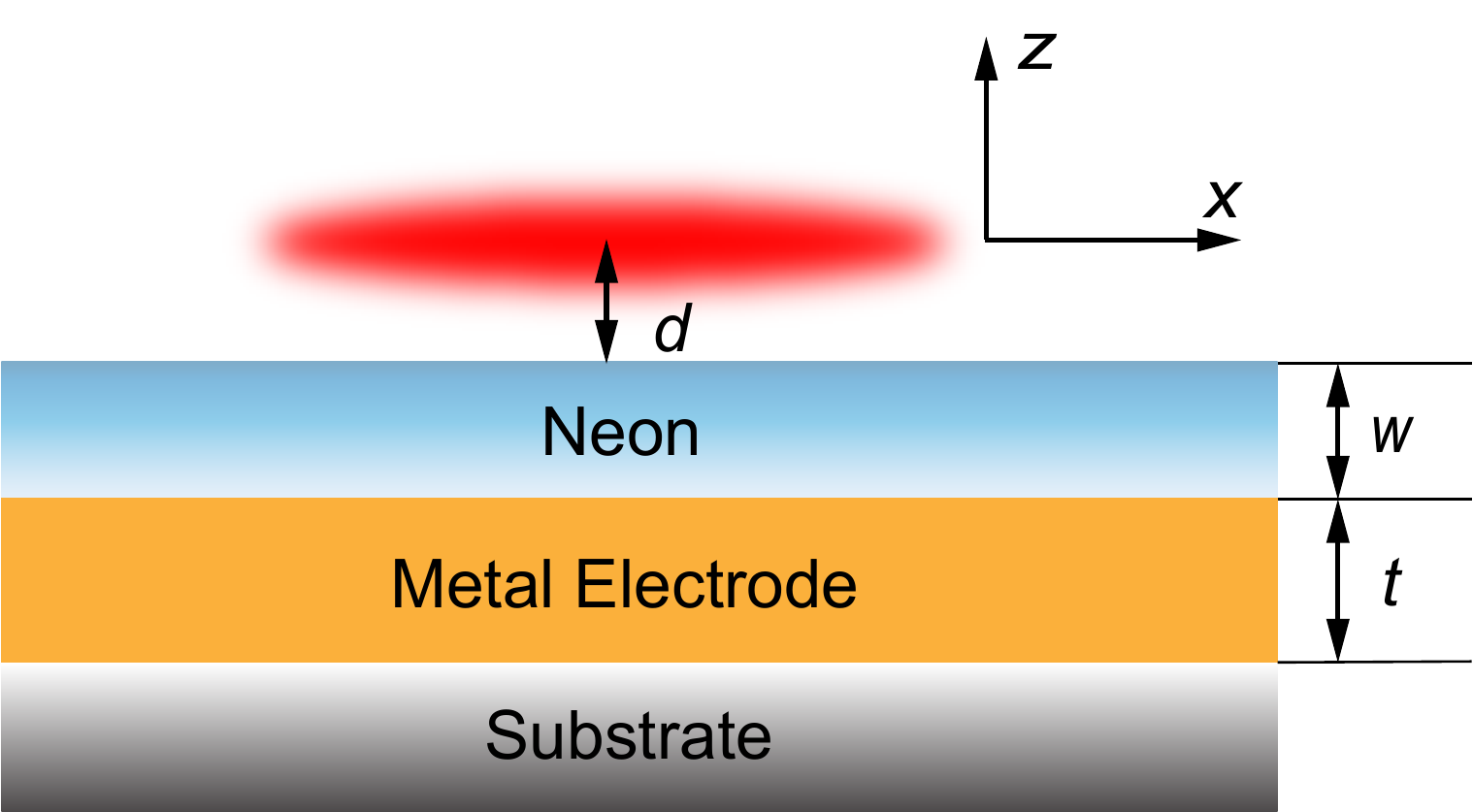}
	\end{center}
	\caption{A schematic diagram of the system configuration. An electron floats in the vacuum above a solid Ne film. Below the solid Ne, there are a metal electrode and a substrate. The thicknesses of solid Ne and metal electrode are $w$ and $t$, respectively. The red cloud represents the electron wavefunction at a mean distance $d\approx1$~nm away from the surface and extending along the surface on the scale of $100$~nm. }
	\label{Fig:1-geometry}
\end{figure}

We consider a system configuration shown in Fig.~\ref{Fig:1-geometry}, which well approximates a real device. A single electron floats above the surface of a solid Ne film of thickness $w$. The solid Ne is grown on top of a metal electrode, usually made of superconducting aluminum (Al) or niobium (Nb), of thickness $t$. Underneath the metal, there is a substrate, usually made of intrinsic silicon (Si) or sapphire (Al$_2$O$_3$) with very low microwave loss. The environmental temperature $T$ can be taken to be 10~mK for a typical circuit quantum electrodynamics (cQED) experiment in a dilution refrigerator. Although the electron's wavefunction mostly resides in vacuum, the solid Ne film, as well as the metal electrode below, can induce spin decoherence.

Natural Ne consists of three stable isotopes: $^{20}$Ne (90.48\%), $^{21}$Ne (0.27\%), and $^{22}$Ne (9.25\%) with the abundance of each component given in the parentheses. $^{20}$Ne and $^{22}$Ne have 0 nuclear spin while $^{21}$Ne has $\frac{3}{2}$ nuclear spin~\cite{hubbs1956}. All Ne atoms in the ground state have closed shells and fully paired electrons. The total angular momentum of the shell electrons is zero and hence does not produce intrinsic magnetic moment. In this section, we systematically investigate various spin decoherence mechanisms and calculate the respective coherence times. We first assume $^{21}$Ne atoms have all been removed, leaving the analysis of their influence for later.

\subsection{Spontaneous emission in vacuum and in a superconducting cavity}

An electron spin qubit is usually operated under an applied constant external magnetic field $B_0$, which gives rise to a Zeeman energy splitting between the up- and down-spin states,
\begin{equation}
\Delta E_{\text{Zeeman}} = g_{\text{s}}\muB B_0 \equiv \hbar\omega \equiv hf.
\end{equation}
Here, $g_{\text{s}} = 2$ is the $g$-factor of free-electron spin, $\muB=e\hbar/2m_{\text{e}}$ is the Bohr magneton (in SI units) with $m_{\text{e}}$ being the free-electron mass and $f=\omega/2\pi$ is the associated Zeeman transition frequency proportional to $B_0$. Their ratio $f/B_0$ is 28~GHz/T. In the most common electron paramagnetic resonance (EPR) studies, the magnetic field is chosen around $B_0\approx0.35$~T so that $f\approx 9.8$~GHz lies in the convenient microwave X band.

If the electron spin is initially in the excited state, the relaxation rate towards the ground state through the spontaneous magnetic-dipole radiation in vacuum is given by~\cite{schweiger2001}
\begin{equation}
W=\frac{2}{3}\frac{\mu_{0}g_{\text{s}}^{2}\muB^{2}}{4\pi \hbar} \frac{\omega^{3}}{c^{3}} \equiv \frac{1}{T_1},
\label{spontaneous emission rate}
\end{equation}
which defines the associated relaxation time $T_1$. For $B_0 = 0.35$~T, there is $T_1 = 5.3\times 10^{11}$~s. This sets an upper bound for the coherence time $T_2$ by the relation~\cite{traficante1991},
\begin{equation}
T_{2}\leq 2T_{1} = 1.0\times 10^{12}~\text{s},
\end{equation}
which is extremely long.

On the other hand, if the electron is placed in a superconducting cavity, as is common in the cQED architecture, the spontaneous emission rate will be modified by the changed electromagnetic environment due to the cavity. This phenomenon is referred as the \emph{Purcell effect}~\cite{baranov2017}. The Purcell factor $F_{\text{P}}$ characterizes the ratio between the spontaneous emission rate of an electron in the cavity to that in vacuum. For a single-mode resonant cavity, it reads~\cite{purcell1946}
\begin{equation}
F_{\text{P}}=\frac{3}{4\pi^{2}}\lambda^{3}\frac{Q}{V},
\label{Purcell factor}
\end{equation}
where $\lambda$ is the wavelength in the cavity, and $Q$ and $V$ are the quality factor and mode volume of the cavity, respectively. Taking a typical superconducting microwave cavity with $Q\sim10^{4}$ and $V\approx10^{-5}\lambda^{3}$~\cite{blais2004}, the Purcell factor is $F_{\text{P}}=7.6\times10^{7}$. As a result, the relaxation time of the electron in the cavity is shortened to $T_1 = 7\times 10^{3}$~s. Subsequently, the coherence time has a upper bound $T_2\leq1.4\times10^{4}$~s, which is much shorter than the upper bound in vacuum but still longer than most other decoherence processes.

\subsection{Phonon induced fluctuation of the Ne diamagnetic susceptibility}

The magnetic response of a Ne atom is to the leading order diamagnetic and is a quantum mechanical effect. The induced magnetization energy is proportional to the square of applied magnetic field and always increases with the field strength irrespective of the field direction~\cite{ashcroft1976}. In our case, both the electron and solid Ne experience a constant external magnetic field $B_{0}=0.35$~T. To be compatible with the superconducting devices, this field should be applied along the $x$ direction that is parallel to the superconducting films and solid Ne surface. It magnetizes the Ne sample through the diamagnetism of Ne atoms. The induced magnetization generates a magnetization surface current. The magnetization current then generates a magnetic field that acts on the spin of electron. During this process, the thermal fluctuations of bulk phonon modes in the solid Ne change the Ne mass density and consequently change the volume magnetic susceptibility. This temporal variation leads to a fluctuating magnetization current and thus a fluctuating magnetic field that acts on the electron. This mechanism induces spin relaxation and decoherence~\cite{Dykman2022}.

The systematic elaboration of this mechanism and detailed derivation of  $T_{1}$ and $T_{2}$ will be given in a separate paper~\cite{Dykman2022}. The derived electron spin relaxation rate is
\begin{equation}
W= \frac{\chi ^{2}\mu _{\text{B}}^{2}B_{0}^{2}\omega}{64\hbar \pi \rho
v_{\text{s}}^{3}d^{2}}\frac{1}{1-e^{-\hbar \omega/k_{\text{B}}T}},  \label{relaxation rate main text}
\end{equation}%
where $\chi =-6.74\times 10^{-5}$ is the volume magnetic susceptibility of solid Ne, $f=\omega/2\pi =2\mu _{\text{B}}B_{0}/h =9.8$~GHz is the associated Zeeman transition frequency, $\rho =m_{\text{Ne}}n$ is the Ne mass density with $m_{\text{Ne}}=3.35\times 10^{-26}$~kg being the atomic mass and $n=4.5\times 10^{28}$~m$^{-3}$ being the particle number density, $v_{\text{s}}=2k_{\text{B}}T_{\text{D}}/(h\left( 6n/\pi \right) ^{1/3})=704$~m/s is the sound velocity in solid Ne with $T_{\text{D}}=74.6$~K being the Debye temperature~\cite{tari2003}, and $d$ is the distance of the electron from the surface, which can be taken as the $z$-position expectation value of the ground-state wavefunction $\langle z\rangle\approx1$~nm~\cite{zhou2022}. To the first order in $\chi $, the pure dephasing rate $\mathit{\Gamma}=T_\varphi^{-1}$ can be found to be exactly zero~\cite{Dykman2022}.

Therefore, to the leading order in $\chi $, the spin relaxation time of the electron is
\begin{equation}
T_{1}=\frac{1}{W}=3.77\times 10^{6}~\text{s}.  \label{t1}
\end{equation}
The coherence time is completely limited by the relaxation process rather than pure dephasing and so can be obtained as
\begin{equation}
T_{2}=2T_1=7.54\times 10^{6}~\text{s}.  \label{T2}
\end{equation}%
Other than the decoherence from the fluctuating magnetization surface current, the time-varying induced magnetization also induces a fluctuating magnetization bulk current $\delta\bm{J}_{m}=\nabla\times\delta\bm{M}$, which generates a fluctuating magnetic field that can limit the coherence of the electron spin. However, we find that this bulk contribution is two orders of magnitude smaller than the surface contribution above~\cite{Dykman2022}.

In addition to the external magnetic field, the electron's generated dipolar magnetic field from its own spin can also magnetize the Ne sample. However, this is a much smaller effect. Consider the electron to be localized at the position $(0,0,d)$. We can find the magnitude of the generated dipolar magnetic field on the nearest Ne atom at $(0,0,0)$ on the surface to be $B_\text{e}=\mu _{0}\mu_{\text{B}}/(2\pi d^3)\approx2$~mT, which is much less than the external $B_{0}$ of our interest. Further, since the electron magnetic dipolar field decays as $d^{-3}$, bulk Ne atoms see a much weaker field than surface Ne atoms do. Therefore, the electron self-induced decoherence through fluctuating diamagnetism can be safely neglected.

\subsection{Thermal current and local spin density fluctuations in metallic electrodes}

In experiments, solid Ne film is usually grown on a metallic electrode. The distance between the electron and the electrode, \ie, the thickness of Ne film, is only several nanometers to hundreds of nanometers. If the electrode is made of normal metals, the electron spin decoherence can arise from thermal current fluctuations in the electrode, which turn into magnetic field fluctuations, known as thermal magnetic noise.

The spin coherence time on a conducting surface in the presence of thermal magnetic noise is given by~\cite{sidles2003}
\begin{equation}
T_{2}=\frac{64\pi w\left(w+t\right) }{\gamma _{\text{e}}^{2}\mu _{0}^{2}\mathrm{Re}\left(\sigma\right)t }\frac{1}{k_{\text{B}}T+\frac{3}{4}\hbar \omega \coth \left(
\frac{\hbar \omega}{2k_{\text{B}}T}\right) },  \label{coherence time}
\end{equation}%
where $\gamma _{\text{e}}$ is the gyromagnetic ratio of the electron, $\sigma$ is the complex electrical conductivity of the metal, $\omega$ is the Zeeman angular frequency for an applied magnetic field $B_{0}$ parallel to the surface, $t$ is the thickness of the metal, and $w$ is the distance between the electron and electrode (refer to Fig.~\ref{Fig:1-geometry}). A higher conductivity $\sigma$ or a smaller distance $w$ gives more prominent decoherence and thus a shorter $T_2$.

Assuming a 100~nm thick high-conductivity metal, such as copper with the resistivity $\approx 2~\text{n} \Omega \cdot $~cm at 10~mK~\cite{matula1979}, and a 10~nm thick solid Ne, we can estimate the coherence time when $B_0=0.35$~T to be $T_{2}=0.18 $~ms. This is much shorter than the $T_2$ from the preceding mechanisms. In Table~\ref{tab:table1}, we list how $T_2$ varies with the thicknesses of metal electrode and Ne film. In all cases, $T_2$ is longer than 0.1~ms, still longer than the $T_2$ in most semiconductor spin qubits~\cite{stano2021}.

\begin{table}[b]
\caption{\label{tab:table1}%
Calculated coherence time $T_{2}$ for a copper electrode at $T=10$~mK (resistivity $\approx 2~\text{n} \Omega \cdot $~cm) with varied thicknesses $t$ of the metal electrode and $w$ of the Ne film.}

\begin{ruledtabular}
\begin{tabular}{llll}
\textrm{$t$ (nm)}&
\textrm{$w$ (nm)}&
\textrm{$T_{2}$ (s)}\\
\colrule
10 & 10 & $3.0\times10^{-4}$ \\
10 & 100 & $1.9\times10^{-2}$ \\
10 & 1000 & 1.7 \\
100 & 10 & $1.8\times10^{-4}$\\
100 & 100 & $3.4\times10^{-3}$ \\
100 & 1000 & 0.18\\
1000 & 10 & $1.7\times10^{-4}$\\
1000 & 100 & $1.8\times10^{-3}$\\
1000 & 1000 & 0.03\\
\end{tabular}
\end{ruledtabular}
\end{table}

Nonetheless, most electrodes are made of superconducting Al or Nb films today~\cite{zhou2022}. Then the dissipation through thermal currents vanishes. The fluctuation also vanishes according to the fluctuation-dissipation theorem. Thus theoretically, the thermal magnetic noise from a superconducting film is zero. Experimentally, thermal magnetic noise in conductors below $4.2$~K was measured using a quantum interference magnetometer~\cite{harding1967}. It was discovered that the noise decreases by more than an order of magnitude when the specimen is superconducting.

The electrons in the metal also carry spins. Hence in principle, there exist spin density fluctuations besides the thermal current fluctuations. Spin density fluctuations can also produce magnetic field fluctuations acting on the excess electron spin~\cite{lyon2006}. However, if the electrodes are made of superconducting films, then this decoherence mechanism is exponentially suppressed. For s-wave superconductors like Al and Nb, the carriers around the Fermi surface are s-wave Cooper pairs with total spin zero. Quasiparticle excitations, whether by charges or spins, are gapped. The gap constant $\Delta$ corresponds to an excitation frequency around 80~GHz for Al and 700~GHz for Nb~\cite{kittel2004}, which are too high to be relevant in the typical experiments at temperatures around 10~mK and operation frequencies around 10~GHz. So long as the superconducting films are everywhere superconducting with no defects like vortices, there should be no spin density fluctuations that can induce spin decoherence.

\subsection{Electron-nuclear spin-spin interaction}

So far, we have ignored the electron spin decoherence via coupling to the nuclear spins of $^{21}$Ne, which has an abundance of $0.27\%=2700$~ppm in natural Ne. The electron spin $\bm{S}$ and $i$th nuclear spin $\bm{I}_i$ are coupled through the hyperfine interaction~\cite{liu2012},
\begin{equation}
H_{\text{hyper}}=\bm{S\cdot }\sum_{i=1}^N\overleftrightarrow{A_{i}}\cdot \bm{I}_{i}=\bm{S\cdot b},  \label{hyperfine interaction}
\end{equation}
where $N$ is the number of nuclear spins, $\bm{b}\equiv\sum_{i=1}^N\overleftrightarrow{A_{i}}\cdot \bm{I}_{i}$ is an effective field (Overhauser field) and the hyperfine interaction tensor $\overleftrightarrow{A_{i}}$ for the $i$~th nuclear spin takes the form of magnetic dipole-dipole interaction in our system,
\begin{equation}
\overleftrightarrow{A_{i}}=\frac{\mu _{0}}{4\pi }\frac{\gamma _{\text{e}}\gamma _{\text{n}}%
}{r_{i}^{3}}\left( \overleftrightarrow{1}-\frac{3\bm{r}_{i}\bm{r}_{i}%
}{r_{i}^{2}}\right) .  \label{hyperfine interaction tensor}
\end{equation}
Here, $\gamma _{\text{e}}$ and $\gamma _{\text{n}}$ are the gyromagnetic ratios of the electron and $^{21}$Ne nuclear spins, respectively, $\overleftrightarrow{1}$ is the identity tensor and $\bm{r}_{i}$ is the displacement vector of the $i$~th $^{21}$Ne spin from the electron spin.

Random thermal distributions of the nuclear spins give rise to a decoherence channel for the electron spin. In an external magnetic field around $B_0 = 0.35$~T, the Zeeman frequency of a $^{21}$Ne nuclear spin is only around 7.7~MHz, equivalent to 400~$\mu$K. The typical 10~mK experimental temperature is much higher than this and yields significant thermal fluctuations of nuclear spin configurations. The random orientations of nuclear spins then lead to a quasi-statically fluctuating Overhauser field that acts on the electron spin and causes decoherence~\cite{yang2017,liu2012}.

We can estimate the inhomogeneous dephasing time $T_2^*$ due to the thermal fluctuations of the Overhauser field (inhomogeneous broadening). We still take the external magnetic field $B_0 = 0.35$~T in $x$ and the electron Zeeman frequency $f=9.8$~GHz, much larger than the hyperfine interaction strength ($<$~1~MHz) and the $^{21}$Ne nuclear spin resonance frequency $\sim 7.7$~MHz . Therefore, we can take the secular approximation for the hyperfine interaction~\cite{du2009,liu2012,yang2017}, which reduces to
\begin{equation}
H_{\text{hyper}}\approx S^{x}\sum_{i=1}^{N}A_{i}^{xx}I_{i}^{x}.  \label{reduced hyperfine interaction}
\end{equation}
Then the Overhauser field becomes
\begin{equation}
b_{x}=\sum_{i=1}^{N}A_{i}^{xx}I_{i}^{x}.  \label{Overhauser field}
\end{equation}

The thermal distribution of the nuclear spins $I_{i}^{x}$ induces an inhomogeneous broadening of the field,
\begin{eqnarray}
\Delta b_{x} &=&\sqrt{\langle b_{x}^{2}\rangle } \notag \\
&=&\sqrt{%
\sum_{i=1}^{N}(A_{i}^{xx})^{2}\langle I_{i}^{x}I_{i}^{x}\rangle }=\frac{3}{2}\hbar\sqrt{\sum_{i=1}^{N}(A_{i}^{xx})^{2}}  \notag \\
&=& \frac{\mu _{0}}{5\hbar }\mu _{\text{B}}\mu _{\text{n}}
nA_{\text{n}}=8994~{\text{Hz}},
\label{inhomogeneous broadening}
\end{eqnarray}
where $\langle\cdots\rangle$ means the average value of operators in the initial states of nuclear spins $I_{i}^x$, $n$ is the Ne atom number density, $A_{\text{n}} = 0.27\%$ is the natural abundance of $^{21}$Ne, and $\mu_{\text{n}} = 0.66\mu _{\text{N}}$ is the magnetic dipole moment of $^{21}$Ne~\cite{makulski2020}. To get this expression, we have taken an approximation by converting summation into integration in a continuous medium and assuming $^{21}$Ne atoms are uniformly distributed.

Under the further assumption of Gaussian distribution of the random Overhauser field, the inhomogeneous dephasing time of the electron spin reads~\cite{du2009,yang2017}
\begin{equation}
T^{*}_{2}(A_{\text{n}}=0.27\%)= \frac{\sqrt{2} }{\Delta b_{x}}= 0.16~{\text{ms}}.
\label{dephasing time}
\end{equation}
Although this inhomogeneous dephasing time of the electron spin is only on the order of $0.1$~ms, the involved thermal noise from the nuclear spins has very low frequencies, \ie, being quasi-static, and can be completely removed by dynamic decoupling techniques~\cite{du2009,liu2012,yang2017}. This can offer several orders of magnitude longer $T_2$ as shown below.

In addition, the dipole-dipole interaction between $^{21}$Ne nuclei may cause pair-wise nuclear-spin flip-flops and hence dynamical quantum fluctuations of the Overhauser field to the electron spin. This will also contribute to decoherence, though in a minor way~\cite{du2009,liu2012,yang2017}. An accurate evaluation of this effect requires solution of  the quantum many-body dynamics of interacting nuclear spins~\cite{liu2008,liu2009} beyond the scope of this paper.

We now exploit an algebraic expression to estimate the electron spin coherence time $T_2$ measured under the application of Hahn echoes~\cite{kanai2022}. For a dilute nuclear spin bath ($<10^{28}$~m$^{-3}$) in a magnetic field strong enough to validate the secular approximation, a phenomenological expression can give a good estimate on the coherence time without exactly solving the spin Hamiltonian and time evolution. For solid Ne with the natural abundance of $^{21}$Ne, the coherence time can be found as~\cite{kanai2022}
\begin{equation}
T_{2}(A_{\text{n}}=0.27\%)= c\left\vert g_{\text{n}}\right\vert^{-1.6}I_{\text{n}}^{-1.1}(A_{\text{n}}n)^{-1}=30~{\text{ms}},
\label{dephasing time}
\end{equation}
where $c=1.5\times10^{24}$~m$^{-3}$~s is an isotope independent constant, $g_{\text{n}}=-0.44$ is the nuclear spin $g$-factor of $^{21}$Ne~\cite{makulski2020}, and $I_{\text{n}}=3/2$ is the the nuclear spin quantum number of $^{21}$Ne.

Practically, the influence of $^{21}$Ne nuclear spins to the electron spin coherence can be suppressed by isotopic purification. Isotopically purified $^{22}$Ne with only 1~ppm of $^{21}$Ne is commercially available (Cryoin Engineering Ltd.)~\cite{bondarenko2018} For 1~ppm of $^{21}$Ne, the estimated inhomogeneous dephasing time $T^{*}_{2}$ is $0.43$~s, and the coherence time $T_2$ under Hahn echoes can reach $81$~s.

\section{Discussion}

Finally, we propose an experimental scheme to realize the single-electron spin qubits on solid Ne in the cQED architecture. Our scheme follows what have been demonstrated in the semiconductor QD spin qubits~\cite{samkharadzev2018,mi2018,landig2018,hu2012,benito2017}. The electron can be trapped in either a single-quantum-dot (SQD) or a double-quantum-dot (DQD) structure on the solid Ne surface. A uniform, in-plane magnetic field in $x$ can be applied along the surface to produce the Zeeman splitting. A pair of cobalt (Co) micromagnets can be fabricated on the two sides of the electron trap to produce a nonuniform magnetic field, whose field direction is mainly in $z$ and gradient direction is mainly in $x$. This field gradient, typically on the order of 1~T/$\mu$m, can provide a synthetic spin-motion coupling. Since we have experimentally demonstrated the strong coupling between the motional (charge) states of the electron and microwave photons in a superconducting resonator, so long as the spin-motion coupling is strong enough, the spin states can also strongly couple with microwave photons. In this framework, known as the electric dipole spin resonance (EDSR), the electron spin states can be addressed and readout by the electric, rather than magnetic, part of microwave photons. Two electron spins can be entangled with each other by placing two electrons at the two ends of a resonator and letting them exchange virtual photons~\cite{borjans2020,sammak2022}. 

However, it is worth mentioning that the spin qubits achieved in this way, though will have a much longer coherence time than that of charge qubits, may not achieve the intrinsic spin coherence time calculated above, because of the introduced permanent spin-motion coupling and the consequent sensitivity to charge noise. If the spin-motion coupling can be made switchable, {\eg} by a superconducting current that can be turned on and off at will~\cite{schuster2010}, then it is more likely to approach the intrinsic spin coherence time. 

Besides, there is one more caution to take. While the in-plane uniform magnetic field, commonly on the order of 0.3~T, and the out-of-plane nonuniform magnetic field from the micromagnets cannot destroy the superconductivity of Nb thin films as a whole, some magnetic fluxes may penetrate through the thin films and be trapped as vortices. Quasiparticles (bound fermion states) inside the vortex cores can have a reduced gap $\Delta'\sim\Delta^{2}/E_{\text{F}}$ ($E_{\text{F}}$ is the electron Fermi energy in Nb)~\cite{caroli1964} that is much smaller than the original superconducting gap $\Delta$. Excitation of these quasiparticles may cause interstate transitions leading to additional spin relaxation and decoherence to the qubit. In-depth studies of these processes will be left in our future work.

\section{Conclusion}

In summary, we have presented a systematic theoretical study of the decoherence mechanisms of a single electron spin on a solid Ne surface at 10~mK temperature. We find that the relaxation time $T_{1}$ of an electron spin in a magnetic field of $0.35$~T is on the order of $10^{11}$~s in vacuum and $10^{3}$~s in a superconducting cavity with quality factor $Q\sim10^{4}$ and mode volume $V$ around $10^{-5}\lambda^3$. These give the upper bounds $\sim 10^{12}$~s and $\sim 10^{4}$~s for the corresponding coherence time $T_{2}$. We then show that without considering nuclear spins of $^{21}$Ne, the coherence time of an electron spin can exceed $10^{6}$~s under the thermal phonon induced fluctuations of Ne diamagnetic susceptibility. We have also investigated the decoherence processes from the thermal magnetic noise in normal metal electrodes made of Cu. In this case, we have $T_{2}\sim 0.17~\text{ms}-1.7$~s depending on the thicknesses of metal and Ne films. Nonetheless, the electrodes in the cQED architecture are usually made of superconducting Al or Nb, and hence do not carry thermal magnetic noise. For a solid Ne sample containing the natural abundance of $0.27\%=2700$~ppm $^{21}$Ne, we estimate that the inhomogeneous dephasing time $T^{*}_{2}$ due to the electron-nuclear spin-spin interaction is around $0.16$~ms. However, the decoherence induced by $^{21}$Ne nuclear spins can be suppressed by isotopic purification of Ne. A solid Ne with 1~ppm of $^{21}$Ne yields $T_2^* = 0.43$~s. Furthermore, this decoherence source can be eliminated by dynamic decoupling techniques thanks to the quasi-static nature of the thermal fluctuations of nuclear spins. The remaining dynamical quantum noises due to nuclear-nuclear dipole interaction can be mitigated by standard Hahn echo techniques. The coherence time $T_{2}$ can be improved to $30$~ms for natural Ne and $81$~s for purified Ne. These results indicate that single-electron spin qubits on solid Ne can become promising new spin qubits that are superior to most existing semiconductor spin qubits.


\begin{acknowledgments}
This work was performed at the Center for Nanoscale Materials, a U.S. Department of Energy Office of Science User Facility, and supported by the U.S. Department of Energy, Office of Science, under Contract No. DE-AC02-06CH11357. D. J., I. M., and Q. C. acknowledge support from Argonne National Laboratory Directed Research and Development (LDRD) Program. D. J. acknowledges additional support from the Julian Schwinger Foundation (JSF) for Physics Research. L. J. acknowledges support from the Packard Foundation (2020-71479). The authors are grateful to Anthony J. Leggett and Mark I. Dykman for inspiring discussion.
\end{acknowledgments}



\nocite{*}

\bibliography{SpinCoherenceRef}

\end{document}